\documentstyle[12pt]{article}
\makeatletter
\newcommand{\singlespacing}{\let\CS=\@currsize\renewcommand{\baselinestretch}
{1.0}\tiny\CS}
\newcommand{\doublespacing}{\let\CS=\@currsize\renewcommand{\baselinestretch}
{1.5}\tiny\CS}

\begin{document}

\title{Nonlocality without inequality for spin-s system}
\author{Samir Kunkri\thanks{{\bf E-mail:skunkri$\_$r@isical.ac.in}}  \thinspace and Sujit K. Choudhary \thanks{{\bf E-mail:sujit$\_$r@isical.ac.in}}\\
Physics and Applied Mathematics Unit, \\
Indian Statistical Institute, \\
Kolkata 700108 , India}

\maketitle

\vspace{0.5cm}
\begin{center}
{\bf Abstract}
\end{center}

{\small We analyze Hardy's non-locality argument for two spin-s
systems and show that earlier solution in this regard was restricted
due to imposition of some conditions which have no role in the
argument of non-locality. We provide a  compact form of non-locality
condition for two spin-s systems and extend it to $n$ number of
spin-s particles. We also apply more general kind of non-locality
argument
still without inequality, to higher spin system.}\\

\section*{Introduction}

The well known contradiction between quantum theory and local
realism was first revealed  by  Bell's inequality \cite{bell}.
Interestingly Hardy gave a proof of non-locality without using
inequality for two spin-$\frac{1}{2}$ particles \cite{hardy93} and
all pure entangled states excepting  maximally entangled one,
exhibit this kind of  non-locality. This result has been generalized
for many qubits and higher dimensional bi-partite system
\cite{karpra, sghosh98, cereceda04}. Kar showed there is no two
qubits mixed state which satisfy Hardy's non-locality argument for
two observabls on each setting \cite{kar97} and Cabello gave another
argument of Bell's theorem without inequalities for GHZ and W state
\cite{cabello02}. Based on Cabello's logic structure Liang et. al
\cite{liang05} provided an example for two qubit mixed state which
shows non-locality still without inequality.

Hardy's argument has been generalized for two spin-s particles  by
Clifton and Niemann \cite{clifton92}. Based on Clifton and Niemann's
argument, Ghosh et.al \cite{ghosh98} showed that for any two spin-s
particles and two measurement possibilities for each particle, there
are infinitely many states (called "Hardy states") showing Hardy's
non-locality. They also showed that for two spin-s particles and for
given choice of observables, the closure of the set of states
showing Hardy-type non-locality constitute a 2s dimensional subspace
of the Hilbert space
associated with the system.\\
In this paper we show that for the case of two spin s particles the
set of solutions found in Ghosh.et.al.\cite{ghosh98}, is restricted
due to use of some constraints which have no role in the
non-locality argument. In this paper we shall show that abandoning
the redundancy will provide a larger set of solutions whose closure
will constitute a $4s^2$ dimensional subspace instead of $2s$
dimensional subspace \cite{ghosh98}   of the Hilbert space
associated with the system. We also provide   a new argument of
non-locality without inequality for two spin-s particle along the
procedure (based on Cabello's logic structure) followed by Liang
et.al.\cite{liang05} where Hardy's
argument is the extreme case of this more generalized argument. \\

\section*{Hardy's non-locality for two spin-s system}

Let us consider two spin-$s$ ($s = \frac{1}{2}, 1,
\frac{3}{2}.......$) particles A and B. Let $s_a$ and
$s_{a^{\prime}}$ represents spin component of particle A along $a$
and $a^{\prime}$ direction. Similarly $s_b$ and $s_{b^{\prime}}$
represents spin component of particle B along $b$ and $b^{\prime}$
direction. The value of $s_a$, $s_b$, $s_{a^{\prime}}$,
$s_{b^{\prime}}$ runs from $-s$ to $+s$.\\ Following Clifton and
Niemann's procedure a state $|\psi\rangle$ is called a Hardy state
if the following condition are satisfied.
\begin{equation}
P(s_a = s_b = s) = 0
\end{equation}
\begin{equation}
P(s_a + s_{b^{\prime}} \ge 0) = 1
\end{equation}
\begin{equation}
P(s_{a^{\prime}} + s_b \ge 0) = 1
\end{equation}
\begin{equation}
P(s_{a^{\prime}}= s_{b^{\prime}} = -s) = p
\end{equation}
One can easily check that equation $(1)$ - $(4)$ are incompatible
with local realism. Interestingly if we express the above equations
in terms of probability on one dimensional projector, one can easily
see that some  equations have no role in constructing the argument
to show the contradiction with local realism. To make it explicit we
first consider the system of two particles, both having spin-1. For
$s = 1$ the above equations can be written as
\begin{equation}
P(s_a = +1, s_b = +1) = 0
\end{equation}
\begin{equation}
P(s_a = -1, s_{b^{\prime}} = -1) = 0
\end{equation}
\begin{equation}
P(s_a = -1, s_{b^{\prime}} = 0 ) = 0
\end{equation}
\begin{equation}
P(s_a = 0, s_{b^{\prime}} = -1) = 0
\end{equation}
\begin{equation}
P(s_{a^{\prime}} = -1, s_b = -1) = 0
\end{equation}
\begin{equation}
P(s_{a^{\prime}} = -1, s_b = 0) = 0
\end{equation}
\begin{equation}
P(s_{a^{\prime}} = 0, s_b = -1) = 0
\end{equation}
\begin{equation}
P(s_{a^{\prime}} = -1, s_{b^{\prime}} = -1) = p
\end{equation}
To show that the above equations contradict local realism we start
from equation $(12)$. This equation tells that if there is Hidden
Variable Theory (HVT) then  there is some hidden variable states for
which $s_{a^{\prime}} = -1$, $s_{b^{\prime}} = -1$.We consider one
such state denoted by $\lambda$ (say). Now for this state  $\lambda$
equations $(6)$ and $(8)$ tell $s_a = +1$. Similarly for those
$\lambda$ equations $(9)$ and $(10)$ tell $s_b = +1$. Then $P(s_a =
+1, s_b = +1)$ should have been non-zero for all $\lambda$ states which
satisfy $(12)$  but this contradicts  equation $(5)$.\\
One should note that to run the Hardy's argument for $s = 1$, we
have not used equations $(7)$ and $(11)$. Hence  Hardy's state need
not be orthogonal to the projectors $P[|s_a = -1, s_{b^{\prime}} = 0
\rangle ]$ and $P[|s_{a^{\prime}} = 0, s_b = -1) = 0\rangle ]$
appearing in $(7)$ and $(11)$. It has to be orthogonal only  to the
projectors $P[|s_a = +1, s_b = +1\rangle ]$, $P[|s_a = -1,
s_{b^{\prime}} = -1,\rangle ]$, $P[|s_a = 0, s_{b^{\prime}} =
-1\rangle ]$, $P[|s_{a^{\prime}} = -1, s_b = 0
\rangle ]$, and  $P[|s_{a^{\prime}} = 0, s_b = -1\rangle ]$.\\

Discarding all the unnecessary restrictions the generalized Hardy's
argument for two spin-s system takes the following form.
\begin{equation}
P(s_a = s_i, s_b = s_j) = 0
\end{equation}
\begin{equation}
P(s_a \ne s_i, s_{b^{\prime}} = s_l) = 0
\end{equation}
\begin{equation}
P(s_{a^{\prime}} = s_k, s_b \ne s_j ) = 0
\end{equation}
\begin{equation}
P(s_{a^{\prime}}= s_k, s_{b^{\prime}} = s_l) = p
\end{equation}
Where $s_i$, $s_j$, $s_k$ and $s_l$  can take  values from $-s$ to
$+s$. For a fixed value of $s_i$, $s_j$, $s_k$ and $s_l$ each of the
sets $(14)$ and $(15)$ contains $2s$ equations. So the total number
of equations in $(13)$, $(14)$ and $(15)$ is $(4s + 1)$.\\ To
satisfy these $(4s + 1)$ equations , Hardy's state has to be
orthogonal to the $(4s + 1)$ linearly independent vectors
corresponding to the projections appearing in equations $(13)$,
$(14)$ and $(15)$. Let $M$ be the closed subspace generated by these
$(4s + 1)$ vectors and let $\overline{M}$ be the orthogonal
complement to $M$. The dimension of $\overline{M}$ is obviously
$4s^2$. Let $M^{\prime}$ be the closed subspace generated by the
above $(4s + 1)$ vectors together with the vector corresponding to
the projection operators appearing in equation $(16)$, and let
$\overline{M^{\prime}}$ be the orthogonal complement to
$M^{\prime}$. Obviously $\overline{M^{\prime}}$ is of dimension
$(4s^2 - 1)$. Then any member of the subset $\overline{M} -
\overline{M^{\prime}}$ (of the Hilbert space associated with the
system), whose closure is of the dimension $4s^2$, is a solution
of equation $(13)-(16)$.\\

\section*{Generalization to $n$ spin-s particles}

The above non-locality argument can easily be extended to $n$ ($n
\ge 2$) number of spin-$s$ system. Let $s_{a_k}$ and
$s_{a^{\prime}_k}$ represent spin component of k th particle along
the vector $a$ and $a^{\prime}$ respectively. The value of $s_{a_k}$
and $s_{a^{\prime}_k}$ runs from $-s$ to $+s$. Then Hardy's
non-locality conditions  for $n$ ($n \ge 2$) number of spin-$s$
system is given by
\begin{equation}
P(s_{a_1} = s^1_i, s_{a_2} = s^2_i, ......... s_{a_n} = s^n_i) = 0
\end{equation}
\begin{equation}
P(s_{a_1} \ne s^1_i, s_{a^{\prime}_2} = s^2_j, ..........
s_{a^{\prime}_n} = s^n_j) = 0
\end{equation}
\begin{equation}
P(s_{a^{\prime}_1} = s^1_j, s_{a_2} \ne s^2_i,  s_{a^{\prime}_3} =
s^3_j.......... s_{a^{\prime}_n} = s^n_j) = 0
\end{equation}
$$........................................$$
\begin{equation}
P( s_{a^{\prime}_1} = s^1_j, s_{a^{\prime}_2} = s^2_j,..........
s_{a^{\prime}_{(n-1)}} = s^{(n-1)}_j, s_{a_n} \ne s^n_i) = 0
\end{equation}
\begin{equation}
P( s_{a^{\prime}_1} = s^1_j, s_{a^{\prime}_2} = s^2_j,..........
s_{a^{\prime}_{n}} = s^n_j) = p
\end{equation}
Where $s^n_i$ and $s^n_j$ can take any value from $-s$ to $+s$. We
have already  checked the case where $n = 2$. Similar kind of
argument will show that equations $(17)-(21)$ contradict local
realism. The states which satisfy above equations has to be
orthogonal to $(2ns + 1)$ linearly independent vectors corresponding
to the projection operators appearing in equations $(17)-(20)$. Let
$M$ be the closed subspace generated by these $(2ns + 1)$ vectors
and let $\overline{M}$ be the orthogonal complement to $M$. The
dimension of $\overline{M}$ is obviously $[(2s + 1)^n - (2ns +1)]$.
Let $M^{\prime}$ be the closed subspace generated by the above $(2ns
+ 1)$ vectors together with the vector corresponding to the
projection operators appearing in equation $(21)$, and let
$\overline{M^{\prime}}$ be the orthogonal complement to
$M^{\prime}$. Obviously $\overline{M^{\prime}}$ is of dimension
$[(2s + 1)^n - (2ns +2)]$. Then any member of the subset
$\overline{M} - \overline{M^{\prime}}$ (of the Hilbert space
associated with the system), whose closure is of the dimension $[(2s
+ 1)^n - (2ns +1)]$, is a solution
of equation $(17)-(21)$.\\

\section*{Cabello-type nonlocality argument for two spin-s system}

The logic structure of Hardy's nonlocality goes as follows: A and B
sometimes happen, A always implies D, B always implies C, but C and
D never happen. Cabello gave another logic structure to prove the
Bell's theorem without inequality for GHZ and W state
\cite{cabello02}. Cabello's logic structure is like that: A and B
sometimes happen, A always implies D, B always implies C, but C and
D happen with lower probability than A and B. Here we will give a
cabello like non-locality argument for two spin-s system. Let us
first consider the case for $s = 1$. We rewrite the equations
$(13)-(16)$ with $q$ (where $0 < q < p$) on the right hand side of
equation $(13)$ and where for $s_i = s_j = +1$ and $s_k = s_l = -1$
\begin{equation}
P(s_a = +1, s_b = +1) = q
\end{equation}
\begin{equation}
P(s_a = -1, s_{b^{\prime}} = -1) = 0
\end{equation}
\begin{equation}
P(s_a = 0, s_{b^{\prime}} = -1) = 0
\end{equation}
\begin{equation}
P(s_{a^{\prime}} = -1, s_b = -1) = 0
\end{equation}
\begin{equation}
P(s_{a^{\prime}} = -1, s_b = 0) = 0
\end{equation}
\begin{equation}
P(s_{a^{\prime}} = -1, s_{b^{\prime}} = -1) = p
\end{equation}
One can check that the above equations contradict local realism when
$q < p$. To show that contradiction, we consider the hidden variable
states $\lambda$ for which $s_{a^{\prime}} = -1$, $s_{b^{\prime}} =
-1$. Now for those states $\lambda$ equations $(25)$ and $(26)$ tell
that the value of $s_b$ must be $+1$. Again for these $\lambda$
equations $(23)$ and $(24)$ tell that the value of $s_a$ must be
$+1$. So $P(s_a = +1, s_b = +1)$ should be at least $p$, which
contradict equation $(22)$ as $q < p$. Instead of putting $q$ (
where $q < p$) in equation $(22)$, one can put $q$ in any one of the
equations $(22)$-$(26)$ and see the contradiction with local
realism in the same way. \\
In the straight forward way one can generalize the above Cabello
like argument for two spin-s system as
\begin{equation}
P(s_a = s_i, s_b = s_j) = q
\end{equation}
\begin{equation}
P(s_a \ne s_i, s_{b^{\prime}} = s_l) = 0
\end{equation}
\begin{equation}
P(s_{a^{\prime}} = s_k, s_b \ne s_j ) = 0
\end{equation}
\begin{equation}
P(s_{a^{\prime}}= s_k, s_{b^{\prime}} = s_l) = p
\end{equation}
where $s_i$, $s_j$, $s_k$ and $s_l$ can take values from $-s$ to
$+s$ and $q < p$. Argument will run in the same way to show
contradiction with local realism and things will remain same in
which ever equations of $(28)$ to $(30)$ we put $q$ on the right hand side.\\
States which satisfy above equations $(28)-(31)$ has to be
orthogonal to the $4s$ linearly independent vectors corresponding to
the projection operator appearing in equations $(29)$ - $(30)$ and
non-orthogonal to the projection operator $P[|s_a = s_i, s_b =
s_j\rangle]$ and $P[|s_{a^{\prime}}= s_k, s_{b^{\prime}} =
s_l\rangle]$. Let $M$ be the closed subspace generated by these $4s$
vectors and let $\overline{M}$ be the orthogonal complement to $M$.
The dimension of $\overline{M}$ is obviously $4s^2 + 1$. The state
vector in $\overline{M}$ must be orthogonal to the above $4s$
linearly independent vectors corresponding to the projection
operator appearing in equations $(29)$ and $(30)$ and non-orthogonal
to the projection operator $P[|s_a = s_i, s_b = s_j\rangle]$ and
$P[|s_{a^{\prime}}= s_k, s_{b^{\prime}} = s_l\rangle]$. Because the
dimension of $\overline{M}$ is $4s^2 + 1$ which is greater than 1,
there are infinitely many vectors satisfying equations $(28)-(31)$.
Then mixture of them also satisfies equations $(28)-(31)$ which
shows Cabello like nonlocality. Here it is interesting to note that
for $s = 1/2$, where standard  Hardy's non-locality argument for
given choice of observables  offers a unique state as solution
\cite{kar97}, there are more than one vector as solution. So there
are mixed states even for two qubits which exhibit this kind of
non-locality \cite{liang05}. In straight forward way one can extend
Cabello like nonlocality argument for $n$ ($n \ge 2$) number of
spin-$s$
system.\\

\section*{Conclusion}

Hardy's version of Bell's theorem has been considered to be
interesting as it does not use inequality explicitly \cite{mermin}.
In higher dimension, construction of Bell's inequality remains a
tough job barring some results \cite{collins}. In this context
various kind of Hardy like argument can be constructed easily in
higher dimensional system which may be very useful to reveal
non-locality of non-trivial density matrix in higher dimension. Here
we reveal the unnecessary restriction of the previous version of
Hardy's non-locality in higher dimension and show that the solution
set is larger than one previously found. We also discuss various
generalization in the context of  new forms of non-locality argument
as well as various cases where more than two particles are involved.

\section*{Acknowledgement}

The authors would like to thank Guruprasad Kar, Debasis Sarkar and
Indrani Chattopadhyay for interesting discussions. S.K acknowledges
the support by the Council of Scientific and Industrial Research,
Government of India, New Delhi.

\end{document}